\documentclass[
showkeys,
preprintnumbers,amsmath,amssymb]{revtex4}

\usepackage{graphicx}
\usepackage{dcolumn}
\usepackage{bm}

\usepackage{graphicx}
\usepackage{epsfig}
\usepackage{color}

\begin{document}

\title{Ermakov-Lewis Invariants and Reid Systems}

\author{Stefan C. Mancas }
\email{stefan.mancas@erau.edu}

\affiliation{Department of Mathematics, Embry-Riddle Aeronautical University,\\ Daytona Beach, FL. 32114-3900, U.S.A.}

\author{Haret C. Rosu}
\email{hcr@ipicyt.edu.mx}
\affiliation{IPICyT, Instituto Potosino de Investigacion Cientifica y Tecnologica,\\
Camino a la presa San Jos\'e 2055, Col. Lomas 4a Secci\'on, 78216 San Luis Potos\'{\i}, S.L.P., Mexico}

\begin{abstract}Reid's $m$th-order generalized Ermakov systems of nonlinear coupling constant $\alpha$ are equivalent to an integrable Emden-Fowler equation. The standard Ermakov-Lewis invariant is discussed from this perspective, and a closed formula for the invariant is obtained for
the higher-order Reid systems ($m\geq 3$). We also discuss the parametric solutions of these systems of equations through the integration of the Emden-Fowler equation and present an example of a dynamical system for which the invariant is equivalent to the total energy.
\end{abstract}

\keywords{Ermakov-Lewis invariant, Reid system, Emden-Fowler equation, Abel equation, parametric solution.\\
Highlights: Reid systems of order $m$ are connected to Emden–Fowler equations. General expressions for the Ermakov-Lewis invariants both for $m=2$ and $m\geq 3$ are obtained. Parametric solutions of the Emden-Fowler equations related to Reid systems are obtained.}

\centerline{Phys. Lett. A 378 (2014) 2113-2117}
\centerline{{\small arXiv:1402.4402}}

\maketitle
For the large class of parametric oscillators (with time-dependent frequencies) and their vast applications, the importance of the Ermakov-Pinney equation with inverse cubic nonlinearity as a helpful auxiliary equation is well established in the literature. The first works in this area have been published in Danish \cite{Steen} and Russian \cite{E80} by two mathematicians of the 19th century and available in English only since the beginning of the 21st century. In 1880, Ermakov discussed the following pair of equations (subindices of one and two letters denote first and second-order derivatives with respect to the independent variable, unless otherwise specified):
\begin{eqnarray}\label{i-1}
 \left\{ \begin{array}{ll}
 q_{tt}+\omega^2(t) q=0~,\\
\tilde{q}_{tt}+\omega^2(t) \tilde{q}= \alpha \tilde{q}^{-3}~. 
\end{array} \right.
\end{eqnarray}
The interesting fact concerning \eqref{i-1} is that the dynamical systems with equations of motion given by the linear equation, and therefore of Hamiltonian $H=\frac{1}{2}\left(p^2+\omega^2(t)q^2\right)$, where $p=q_t$, are further endowed with the so-called Ermakov-Lewis (EL) invariant which depends on the nonlinear constant $\alpha$ and is constructed from any solutions $q$ and $\tilde{q}$ of \eqref{i-1} as follows:
\begin{equation}\label{i-2}
I=\alpha\left(\frac{q}{\tilde{q}}\right)^2+(\tilde{q}q_t-q\tilde{q}_t)^2~.
\end{equation}
If $\alpha=0$ then $I=W^2$, where $W$ is the Wronskian of two linearly independent solutions of the linear equation. Thus, mathematically, the invariant $I$ is closely related to the Wronskian. Modern research related to this invariant started in the second half of the 1960s \cite{lewis67} when Lewis rediscovered it in a completely different theoretical framework and also provided the first application in quantum mechanics upon turning it into a Hermitian dynamical operator by considering $q$ and $p$ as operators but keeping $\tilde{q}$ as a $c$-number and showing that it was a constant of motion and thus possessing time-independent eigenvalues. For a recent general discussion of dynamical invariants in the quantum-mechanical framework we recommend Section 2 in \cite{lohe09}. In the following, we focus only on the classical aspects of the problem. For one-dimensional time-dependent classical Hamiltonians with more general potentials other than $\frac 1 2 \omega^2(t)q^2$, we mention the general result of Lewis and Leach who obtained all potentials that admit an invariant quadratic in $p$ and who determined all those invariants \cite{ll82}. Besides, they considered the possibility to find more general invariants that are polynomials in $p$ of higher degree than quadratic.

\medskip

Much less attention has been paid in the literature to the higher degree nonlinear generalization of Ermakov systems of equations introduced by Reid in 1971 \cite{reid}:
\begin{eqnarray}
\left\{ \begin{array}{ll}
q_{tt}+\omega^2(t) q&=0 \label{sys1}\\
\tilde{q}_{tt}+\omega^2(t) \tilde{q}&=\alpha (q_1q_2)^{m-2}\tilde{q}^{1-2m}~, \qquad \alpha \neq 0~, \qquad m \ne 0,1, \quad m \in \mathbb{N}~,\label{sys2} \end{array}\right. 
\end{eqnarray}
where $q_1$ and $q_2$ are particular linear independent solutions of the homogeneous linear equation of system \eqref{sys1}.
The standard Ermakov systems are included as the particular case $m=2$.

Reid has shown that the following nonlinear superposition
\begin{equation}\label{ve}
\tilde{q}(t)=\Big(q_1^m+\frac{\alpha}{(m-1)W^2}q_2^m\Big)^{\frac 1 m}
\end{equation}
is a solution to the nonlinear equation in (\ref{sys2}). 
In the particular case $m=2$, Reid's formula \eqref{ve} reduces to Pinney's formula corresponding to Ermakov's systems \cite{Pinney}
\begin{equation}\label{ve1}
\tilde{q}_{_{\rm Pin}}(t)=\Big(q_1^2+\frac{\alpha}{W^2}q_2^2\Big)^{\frac 1 2}~.
\end{equation}


In this Letter, our main goal is to get general formulas for the EL invariant both in the standard $m=2$ case and the higher order cases $m>2$
by using the integration of the corresponding Emden-Fowler equations.
For that, we apply to \eqref{sys1} Ermakov's idea \cite{E80} of eliminating $\omega^2(t)$, which leads to:
\begin{equation}\label{3bis}
\frac{d}{d t}(q\tilde{q}_{t}-\tilde{q}q_{t})=\alpha (q_1q_2)^{m-2}\tilde{q}^{1-2m}q~.
\end{equation}
Multiplying both sides by $q\tilde{q}_{t}-\tilde{q}q_{t}=-\tilde{q}^2\frac{d}{d t}\big(\frac{q}{\tilde{q}} \big)$ we get
\begin{equation}\label{4}
\frac{d}{d t}(q\tilde{q}_{t}-\tilde{q}q_{t})^2=-\alpha (q_1q_2)^{m-2}\tilde{q}^{4-2m}\frac{d}{d t}\Big(\frac{q}{\tilde{q}} \Big)^2~.
\end{equation}
Since $q_1$ and $q_2$ are particular solutions of Wronskian $W$, then let us use
\begin{equation}\label{6}
q_2=Wq_1\int \frac{1}{q_1^2} dt
\end{equation}
which is  the reduction of order formula. Substituting this into \eqref{4} it yields
\begin{equation}\label{7bis}
\frac{d}{d t}(q\tilde{q}_{t}-\tilde{q}q_{t})^2=-\frac{\alpha W^{m-2}}{m-1}\Big(\int \frac{d t}{q^2}\Big)^{m-2} \frac{d}{d t}\Big(\frac {q^2} {\tilde{q}^2} \Big)^{m-1},
\end{equation}
where the subindex of $q_1$ has been dropped.  

Let us introduce a transformation defined by
\begin{align}
\tilde{r}&=\frac{\tilde{q}}{q}~,\\
Y&=\int \frac {d t}{q^2}
\end{align}
and notice that the Wronskian built from the solutions $q$ and $\tilde{q}$ is the $\tilde{r}_Y$ derivative:
\begin{equation}\label{8}
q\tilde{q}_{t}-\tilde{q}q_{t}=q^2 \frac{d\tilde{r}}{dt}=\frac{d\tilde{r}}{dY}=\tilde{r}_Y.
\end{equation}
This is a very useful relationship that helps us to obtain the corresponding Emden-Fowler equation. When we substitute all the above into \eqref{7bis} we obtain
 \begin{equation}\label{9}
\frac{d}{d t}\Big(\frac {d\tilde{r}}{dY}\Big)^2=-\frac{\alpha W^{m-2}}{m-1}Y^{m-2}\frac{d}{d t}\tilde{r}^{2-2m}.
\end{equation}
Now, we multiply both sides of \eqref{9} by $\tilde{r}^2=\frac{d t}{d Y}$ and get
 \begin{equation}\label{10}
\frac{d}{d Y}\Big(\frac {d\tilde{r}}{dY}\Big)^2=-\frac{\alpha W^{m-2}}{m-1}Y^{m-2}\frac{d}{d Y}\tilde{r}^{2-2m},
\end{equation}
which simplifies to the following Emden-Fowler (EF) equation
 \begin{equation}\label{11}
\tilde{r}_{YY}=\alpha W^{m-2}Y^{m-2}\tilde{r}^{1-2m}
\end{equation}
that can be considered as equivalent to the initial Reid system of order $m$.

According to Polyanin \cite{Pol}, a particular solution  is
 \begin{equation}\label{13}
\tilde{r}_{_{\rm Pol}}=(-4 \alpha W^{m-2})^{\frac {1}{2m}} \sqrt{Y}~.
\end{equation}
The integrability of the EF equation \eqref{11} resides in the particular powers of $\tilde{r}$ and $Y$, and the general solution can be written in parametric form. The fact that Reid's systems are integrable plays an important role in the integrability of \eqref{11}.

\bigskip

\noindent To proceed further, we separate the well-studied case $m=2$ from the higher-order cases $m\geq 3$.

\medskip

\noindent {\bf Case 1}. $m=2$. In this case, \eqref{11} reduces to
 \begin{equation}\label{12}
\tilde{r}_{YY}=\alpha \tilde{r}^{-3}.
\end{equation}
To obtain the EL invariant, let us multiply \eqref{12} by $\tilde{r}_Y$ and integrate once to get
 \begin{equation}\label{15}
 \frac{1}{2}(\tilde{r}_Y)^2=-\frac{1}{2}\alpha \tilde{r}^{-2}+C,
\end{equation}
which after identifying $C\equiv I$, where $I$ is the EL invariant, yields
 \begin{equation}\label{16}
I(\tilde{r},\tilde{r}_Y)=\frac{1}{2}\bigg[(\tilde{r}_Y)^2+\alpha \tilde{r}^{-2}\bigg]~.
\end{equation}

Furthermore, it is easy to show that $I$ is a constant given by the following formula
 \begin{equation}\label{16-L}
 I=\frac{1}{2}\left(a^2\alpha+b^2W^2\right)~,
 \end{equation}
where $a$ and $b$ are the superposition constants of the general solution $q$.

Indeed, \eqref{16} can be written as
 \begin{equation}\label{16a}
I(t)=\frac{1}{2}\bigg[(q\tilde{q}_{t}-\tilde{q}q_{t})^2+\alpha \Big(\frac{\tilde{q}}{q} \Big)^{-2}\bigg]~.
\end{equation}
Then, using the general solution $q$ as the linear superposition $q=aq_1+bq_2$ and
\begin{equation}\label {ve2}
\tilde{q}(t)=\sqrt{q_1^2+\frac{\alpha}{W^2}q_2^2}
\end{equation}
in \eqref{16a}, one gets \eqref{16-L}. Thus, given the initial conditions, the constant value of the Wronskian of the two linear independent solutions, and the nonlinearity parameter, the EL invariant can be calculated from the general formula \eqref{16-L}, which, to the best of our knowledge, was not previously mentioned in the literature. Notice that
from the strict mathematical viewpoint this invariant can be zero if the superposition constants are chosen such that $\frac{b}{a}=\pm\frac{1}{W}\sqrt{-\alpha}$. A lemma in the literature states that the EL invariant is positive semidefinite \cite{fg09}.
According to our result this implies the following condition on the nonlinear coupling $\alpha>-(bW/a)^2$.

  \medskip

We turn now to the solutions of equation \eqref{12}. For this, we decode the equation as the nonlinear Ermakov equation in the particular case of zero frequency $\omega(Y)=0$ that leads further to solutions of \eqref{sys2} in known forms. Since \eqref{12} is an Ermakov equation, a particular solution can be written in terms of the two linearly independent solutions $r_1=1$ and $r_2=Y$ of the homogeneous equation $r_{YY}=0$ using the Pinney formula \cite{Pinney}
 \begin{equation}\label{Psol}
\tilde{r}_{_{\rm Pin}}(Y)=\sqrt{r_1^2+\frac{\alpha}{W^2}r_2^2}\equiv\sqrt{1+\alpha Y^2}~.
 \end{equation}
But knowledge of $\tilde{r}(Y)$ implies getting $\tilde{q}$ from
 \begin{equation}\label{Psol1}
 \tilde{q}_{_{\rm Pin}}=q_1\tilde{r}_{_{\rm Pin}}(Y)\equiv\sqrt{q_1^2+\alpha q_2^2}\,~,
 \end{equation}
 which is the typical Pinney formula when $W=1$.

 On the other hand, Polyanin's particular solution \eqref{13} corresponds to:
 \begin{equation}\label{Psol2}
 \tilde{q}_{_{\rm Pol}}=q_1\tilde{r}_{_{\rm Pol}}(Y)=(-\alpha)^{\frac{1}{4}}\sqrt{2q_1q_2}~.
 \end{equation}
Of course, the particular solutions \eqref{Psol1} and \eqref{Psol2} can be obtained from the general solution of \eqref{12}.
Suppose we consider now the solutions $r_1=1$ and $r_2=Y$ of Wronskian $W=1$ of the homogeneous equation $r_{YY}=0$.
Then, it is known that the general Pinney solution of \eqref{12} can be written as follows \cite{CdeL-08,haas-10}:
\begin{equation}\label{Psol3}
\tilde{r}(Y)=\sqrt{\alpha_1+\alpha_2Y^2+2\alpha_3Y}~,
 \end{equation}
with the $\alpha$ constants fulfilling the condition $\alpha_1\alpha_2-\alpha_3^2=\frac{\alpha}{W^2}$. One can easily see that $\tilde{r}_{_{\rm Pin}}(Y)$ and $\tilde{r}_{_{\rm Pol}}(Y)$ are just particular cases of $\tilde{r}(Y)$ in \eqref{Psol3}.

\medskip

\noindent {\bf Case 2}. When $m>2$, to find the invariant we will use two methods.\\


\noindent (i) Using the substitutions 
$\tilde{r}(Y)=\frac{\tilde{Q}(\tau)}{\sqrt{\tau}}$ and $\tau=\frac{1}{Y}$ in equation \eqref{11}, one gets
  \begin{equation}\label{EFsingleNL}
  \tau^2\tilde{Q}_{\tau\tau}+\tau\tilde{Q}_\tau-\frac{1}{4}\tilde{Q}=\alpha W^{m-2}\tilde{Q}^{1-2m}~.
 \end{equation}
 To get rid of the damping term, one can use Euler's exponential change of independent variable $\tau=e^{\eta}$ that leads to
 the following Reid equation of constant frequency $\omega=\frac{i}{2}$
  \begin{equation}\label{Reid-again}
  \tilde{Q}_{\eta\eta}-\frac{1}{4}\tilde{Q}=\alpha W^{m-2}\tilde{Q}^{1-2m}~.
 \end{equation}
Its solution can be written using Reid's formula \eqref{ve} 
\begin{equation}\label{Reid-again-bis}
\tilde{Q}(\eta)=\Big(e^{+\frac{m\eta}{2}}+\frac{\alpha W^{m-2}}{m-1}e^{-\frac{m\eta}{2}}\Big)^{\frac 1 m}~,
\end{equation}
where the exponential functions are the linear independent solutions of Wronskian $W=-1$ of the hyperbolic oscillator equation
  \begin{equation}\label{Reid-again-tris}
  Q_{\eta\eta}-\frac{1}{4}Q=0~.
 \end{equation}
 Multiplying \eqref{Reid-again} by $\tilde{Q}_\eta$ and integrating, one immediately gets
  \begin{equation}\label{Im1}
  \tilde{Q}_\eta^2=\frac{1}{4}\tilde{Q}^2-\frac{\alpha W^{m-2}}{m-1}\tilde{Q}^{2-2m}+2I\equiv P(\tilde{Q})~,
 \end{equation}
 which provides the general expression for the Ermakov-Lewis invariant for $m>2$:
  \begin{equation}\label{Im2}
  I(\tilde{Q}, \tilde{Q}_\eta)= \frac 1 2\bigg[ \tilde{Q}_\eta^2+\frac{\alpha W^{m-2}}{m-1}\tilde{Q}^{2-2m}-\frac{1}{4}\tilde{Q}^2\bigg]~.
  \end{equation}

  \medskip

Equation \eqref{Im2} can be also written as a function of corresponding Reid's $\tilde{r}$'s and $\tilde{r}_Y$'s as follows:
  \begin{equation}\label{Imq}
  I(\tilde{r},\tilde{r}_Y)=\frac 1 2\bigg[Y(\tilde{r}_Y)^2-\tilde{r}_Y\tilde{r}+\frac{\alpha W^{m-2}}{m-1}\left(\frac{Y}{\tilde{r}^2}\right)^{m-1}\bigg]~.
  \end{equation}

  In terms of $q,\tilde{q}$ the above becomes
\begin{equation}\label{invar}
I(t)=\frac 1 2\bigg[(q\tilde{q}_{t}-\tilde{q}q_{t})^2\int \frac{d t}{q^2}-\frac{\tilde{q}}{q} (q\tilde{q}_{t}-\tilde{q}q_{t})+\frac{\alpha W^{m-2}}{m-1}\Big(\frac{q^2}{\tilde{q}^2}\int \frac{dt}{q^2}\Big)^{m-1}\bigg]~,
\end{equation}
which is the higher order equivalent of \eqref{16a}.

If in \eqref{Imq} we now substitute the Polyanin particular solution we find the following constant value for the Reid invariant for all $m\geq 3$:
  \begin{equation}\label{rezultat}
  I=
  -\frac{(-4\alpha W^{m-2})^{\frac{1}{m}}}{8}\frac{m}{m-1}~.
  \end{equation}

     \medskip

     We now address the issue of finding the Reid solution \eqref{ve} from the solution of the EF equation \eqref{11} for $m>2$.
Formula \eqref{Im1} is separable as follows:
      \begin{equation}\label{phase1}
      \int \frac{\pm d\tilde{Q}}{\sqrt{P(\tilde{Q})}}=\int d\eta =\eta-\eta_0=\ln|\tau|-\ln|\tau_0|=\ln|\frac{\tau}{\tau_0}|~,
       \end{equation}
        which allows us to introduce the exponential parametric form of the Emden-Fowler solutions:
         \begin{align}\label{phase3}
         \tau(\tilde{Q})&=|\tau_0|e^{\pm \int^{\tilde{Q}} \Theta(\tilde{Q}') d\tilde{Q}'}~\\
         y(\tilde{Q})&=\frac{\tilde{Q}}{\sqrt{|\tau_0|}}e^{\mp \frac 1 2 \int^{\tilde{Q}} \Theta(\tilde{Q}')d\tilde{Q}'}~,
         \end{align}
         where $\Theta(\tilde{Q})=P^{-\frac 1 2}(\tilde{Q})$.
  In our case the solutions are:
    \begin{align}
         Y(\tilde{Q})&=\frac{1}{|\tau_0|}e^{\mp \int^{\tilde{Q}} \big (\frac{1}{4}\tilde{Q}'^2-\frac{\alpha W^{m-2}}{m-1}\tilde{Q}'^{2-2m}+2I\big)^{-\frac 1 2}  d\tilde{Q}'}~ \label{phase3-1}\\
         \tilde{r}(\tilde{Q})&=\frac{\tilde{Q}}{\sqrt{|\tau_0|}}e^{\mp \frac{1}{2} \int^{\tilde{Q}} \big (\frac{1}{4}\tilde{Q}'^2-\frac{\alpha W^{m-2}}{m-1}\tilde{Q}'^{2-2m}+2I\big)^{-\frac{1}{2}}  d\tilde{Q}'} \label{phase3-2},
         \end{align}
  where $\tau_0$ is a constant of integration. For example, in the $m=3$ case one gets $I=\frac{3}{8}\left(\frac{\alpha W}{2}\right)^{\frac{1}{3}}$, and then the parametric solutions to the EF equation are as follows:
    \begin{align}\label{phase4}
         Y(\tilde{Q})&=\frac{1}{|\tau_0|}e^{\mp \int^{\tilde{Q}} \big (\frac{1}{4}\tilde{Q}'^2-\frac{\alpha W}{2}\tilde{Q}'^{-4}+\frac{3}{4}\left(\frac{\alpha W}{2}\right)^{\frac{1}{3}}\big)^{-\frac 1 2}  d\tilde{Q}'}~\\
         \tilde{r}(\tilde{Q})&=\frac{\tilde{Q}}{\sqrt{|\tau_0|}}e^{\mp \int^{\tilde{Q}} \big (\frac{1}{4}\tilde{Q}'^2-\frac{\alpha W}{2}\tilde{Q}'^{-4}+\frac{3}{4}\left(\frac{\alpha W}{2}\right)^{\frac{1}{3}}\big)^{-\frac 1 2}  d\tilde{Q}'}~.
         \end{align}

   From \eqref{phase3-1} and \eqref{phase3-2} it follows that $\tilde{r}$ and $Y$ for higher order Reid systems are connected by
  \begin{equation}\label{qYR}
  \tilde{r}=\tilde{Q}\sqrt{Y}
   \end{equation}
   and therefore the parametric solutions differ from the particular Polyanin solution through the function $\tilde{Q}$ that replaces the constant factor $(-4 \alpha W^{m-2})^{\frac {1}{2m}}$. Moreover, using $e^\eta=\tau=\frac{1}{Y}$ in \eqref{Reid-again-bis} one gets
    \begin{equation}\label{qYR1}
    \tilde{Q}(Y)=\left(Y^{-\frac{m}{2}}+\frac{\alpha W^{m-2}}{m-1}Y^{\frac{m}{2}}\right)^{\frac{1}{m}}~.
    \end{equation}
    Then \eqref{qYR} gives
     \begin{equation}\label{qYR2}
     \tilde{q}=\tilde{Q}q_1\sqrt{Y}=\tilde{Q}\sqrt{\frac{q_1q_2}{W}}=\left(Y^{-\frac{m}{2}}+\frac{\alpha W^{m-2}}{m-1}Y^{\frac{m}{2}}\right)^{\frac{1}{m}}
     \sqrt{\frac{q_1q_2}{W}}~.
    \end{equation}
    Since $Y=\frac{q_2}{Wq_1}$, we obtain the final result
    \begin{equation}\label{qYR3}
    \tilde{q}=\Big(q_1^m+\frac{\alpha}{(m-1)W^2}q_2^m\Big)^{\frac 1 m}~,
\end{equation}
which is Reid's formula for solution $\tilde{q}$.

  \bigskip


\medskip

\noindent (ii) One can also solve the full Emden-Fowler equation by the reduction of order method using both dependent-independent variable substitutions
\begin{align}\label{19}
z&=\Big(\frac{Y}{\tilde{r}^2}\Big)^m~,\\
u&=Y \frac{\tilde{r}_Y}{\tilde{r}}
\end{align}
that lead to a first order Abel equation
 \begin{equation}\label{20}
\Big(u-\frac 1 2\Big)u_z=\frac {u^2}{2mz}-\frac{u}{2mz}-\frac{\alpha W^{m-2}}{2m}.
\end{equation}
Now, if  we let $u-\frac 1 2=\frac 1 v$ we obtain the Bernoulli equation
 \begin{equation}\label{21}
v_z=-\frac{1}{2mz}v+\frac{1+4\alpha W^{m-2}z}{8mz}v^3~.
\end{equation}
This can be linearized by $\varphi=v^{-2}$ to give
 \begin{equation}\label{22}
\varphi_z-\frac {1}{mz}\varphi=-\frac{\alpha W^{m-2}}{m}-\frac{1}{4mz}~.
\end{equation}
The solution to \eqref{22} is
 \begin{equation}\label{23}
\varphi(z)=\frac{\alpha W^{m-2}}{1-m}z+Iz^{\frac 1 m}+\frac 1 4~,
\end{equation}
where $I$ is the integration constant which is equivalent to the EL invariant.
Now we use back all the substitutions and solve for $I$ to obtain the same as equation \eqref{Imq} obtained previously.\\

\bigskip

\noindent {\em Lagrangian and Hamiltonian functions}. The inverse transformation of independent variable $\tau=\frac 1 Y$  turns the EF equation \eqref{11} to the EF in the normal form
\begin{equation}\label{2}
\tau \ddot{\tilde{r}}+2 \dot{\tilde{r}}=\alpha W^{m-2} \tau^{-m-1}\tilde{r}^{1-2m}~,
\end{equation}
where the overdot notation for the derivative with respect to $\tau$ is used.
Djukic showed that the following Lagrangean \cite{Djukic,Silva}
\begin{equation}\label{3}
L(\tilde{r},\dot{\tilde{r}})=\frac{\tau^{-2}}{2}\left(\tau^4 \dot{\tilde{r}}^2-\frac{\alpha W^{m-2}}{m-1}\tau^{-(m-2)}\tilde{r}^{2-2m}\right)
\end{equation}
generates \eqref{2} from the variational formulation of Euler-Lagrange equations $\frac{d}{d\tau}\left(\frac{\partial L_m}{\partial{\dot {\tilde{r}}}}\right)-\frac{\partial L_m}{\partial \tilde{r}}=0$, and one can build the Hamiltonian from $H(\mathfrak{p},\tilde{r})=\frac{\partial L_m}{\partial {\dot{\tilde{r}}}}\dot{\tilde{r}}-L_m=\mathfrak{p} \dot{\tilde{r}}-L_m$, where $\mathfrak{p}=\tau^2\dot{\tilde{r}}$, which gives
\begin{equation}\label{3a}
H(\mathfrak{p},\tilde{r})=\frac{\tau^{-2}}{2}\left(\mathfrak{p}^2 +\frac{\alpha W^{m-2}}{m-1}\tau^{-(m-2)}\tilde{r}^{2-2m}\right)~.
\end{equation}

In terms of the original variables $\tilde{r}(Y)$  
and $\tilde{r}_Y=-\tau^2 \dot{\tilde{r}}=-\mathfrak{p}$, the Lagrangian and Hamiltonian functions have the more symmetric forms:

\begin{align}\label{LH}
L(\tilde{r},\tilde{r}_Y)&=\frac{Y^2}{2}\left(\tilde{r}_Y^2-\frac{\alpha W^{m-2}}{m-1}Y^{m-2}\tilde{r}^{2-2m}\right)~,\\
H(\mathfrak{p},\tilde{r})&=\frac{Y^2}{2}\left(\mathfrak{p}^2 +\frac{\alpha W^{m-2}}{m-1}Y^{m-2}\tilde{r}^{2-2m}\right)~.
\end{align}

The integral of motion (33) is obtained using the change of dependent variable $\tilde{r}(Y)=
\frac{\tilde{r}(\tau)}{\sqrt \tau}$. In terms of the canonical variables of the Lagrangian $L$, the invariant (33) becomes
\begin{equation}\label{7}
I(\tilde{r},\dot{\tilde{r}})=
\tau^3 \dot{\tilde{r}}^2+\tau^2 \dot{\tilde{r}} \tilde{r} +\frac{\alpha W^{m-2}}{m-1}\tau^{1-m}\tilde{r}^{2-2m}~.
\end{equation}
This form of the EL invariant has been obtained by Djukic, and also generalized by Rosenau \cite{rosenau}, but in different notations and context and without calling it as such.

One can also ask what kind of integral of motion is the higher-order EL invariant. If one writes down the Hamiltonian of the nonlinear Reid oscillator
\begin{equation}\label{hreid}
H_R(\tilde{p},\tilde{q})=\frac 12[\tilde{p}^2+\omega^2(t)\tilde{q}^2+\alpha\frac{(q^{2}WY)^{m-2}}{m-1}\tilde{q}^{2(1-m)}]~,
\end{equation}
one can show that the total time derivative of the invariant as given in \eqref{invar} is zero:
  \begin{equation}\label{ttd}
\frac{d I}{dt}=\frac{\partial I}{\partial t}+\{I,H_R\}=\frac{\partial I}{\partial t}+\frac{\partial I}{\partial \tilde{q}}\frac{\partial H_R}{\partial \tilde{p}}-\frac{\partial I}{\partial\tilde{p}}\frac{\partial H_R}{\partial \tilde{q}}\equiv0~.
\end{equation}
However, the situation is different from the linear parametric oscillators because now both the EL invariant and the Hamiltonian depend on the nonlinear coupling constant. It can be shown, see also the example that follows, that this invariant can be identified, up to a possible scaling, with the Hamiltonian of the nonlinear Reid oscillator.

\bigskip

\noindent {\em A dynamical system for $m>2$}. In the standard Ermakov case, Eliezer and Gray provided the interpretation of the classical EL invariant as the integral of motion of angular momentum for a two-dimensional auxiliary motion in a closed orbit \cite{Eliezer} and generalizations to three dimensions also exist \cite{gl}. On the other hand, Haas constructed Poisson's structures for Ermakov systems using the Ermakov invariant as the Hamiltonian \cite{h02}. In general, it is not easy to find a physical example with nonlinear singularities stronger than the Ermakov inverse cubic one.

Here, we adapt an application from the Kepler classical mechanics that was previously discussed by Nowakowski and Rosu \cite{Ros}. We assume that the equation for the energy conservation with power law radial potential $V(R)=KR^\epsilon$ can be written in the form (henceforth we use the dot notation for the time derivative):
\begin{equation}\label{en}
E=\frac 1 2 M \dot{R}^2+\frac{1}{2(m-1)}\frac{(-1)^{m-2}l^2}{MR^{2(m-1)}}+V(R)=const.
\end{equation}
The case $m=2$, $\epsilon=-1$ is the standard Kepler case for which the first term is the kinetic energy, the second term is the centrifugal barrier, and the third the gravitational potential \cite{Ros}.

\medskip

We take now the time derivative of \eqref{en} to obtain the Reid equation
\begin{equation}\label{re}
\ddot{R}+\frac{1}{M}\frac{d V}{dR}=\frac{(-1)^{m-2}l^2}{M^2}R^{1-2m}
\end{equation}
 and we notice that \eqref{re} is the same as \eqref{Reid-again} if $t=\eta,~ R=\tilde{Q},~ \frac{l^2}{M^2}=\alpha$, and $\frac{1}{M}\frac{d V}{d\tilde{Q}}=-\tilde{Q}/4$, 
 which gives the quadratic potential $V(\tilde{Q})=K\tilde{Q}^2$, with coupling constant $K=-M/8$ corresponding to a hyperbolic oscillator of imaginary frequency $\omega=\frac 12 i$. Thus, this $\epsilon=2$ case is integrable and we use the solution \eqref{Reid-again-bis} to write the particular solution of \eqref{re} as
\begin{equation}\label{r}
R(t)=\Big(e^{\frac{m}{2} t}+\frac{(-1)^{m-2}l^2}{M^2(m-1)}e^{-\frac{m}{2}t}\Big)^{\frac 1 m}~, ~~~W=-1.
\end{equation}
The meaning of the invariant is shown by writing \eqref{Im2} in terms of $R$:
\begin{equation}\label{Imr}
  I(R, \dot{R})= \frac 1 2\bigg[ \dot{R}^2+\frac{(-1)^{m-2}l^2}{M^2(m-1)}R^{2(1-m)}-\frac{1}{4}R^2\bigg]
  \equiv\frac{1}{M}\bigg[\frac 1 2 M\dot{R}^2+\frac{(-1)^{m-2}\alpha}{2(m-1)}MR^{2(1-m)}-\frac1 8MR^2\bigg]~.
  \end{equation}
Thus, using the hyperbolic radial oscillator as dynamical system, the higher-order EL invariant is the total energy per unit of mass, i.e., the sum per unit of mass of the kinetic radial energy, the energy due to the Reid nonlinearity, which is repulsive if $m$ is even and attractive if $m$ is odd, and a third term which can be interpreted as the potential energy.

\medskip

In conclusion, in this Letter we derived expressions for the Ermakov-Lewis invariant in the case of the higher-order Reid generalization of the Ermakov systems of equations. We also provided an example of a dynamical system for which this invariant is essentially its total energy per unit of mass. We finally mention that similarly to the case of the standard Ermakov-Lewis invariant \cite{Ray81,qin}, another usage of the higher-order invariants is to obtain exact solutions of time-dependent problems \cite{mr2}.

\bigskip
\bigskip

\noindent Acknowledgments: We wish to thank the referees for very useful remarks.

\end{document}